\documentclass[aps,prb,twocolumn,superscriptaddress]{revtex4}
\usepackage{graphicx,latexsym}

\begin{document}

\title{Local magnetization fluctuations in superconducting glasses resolved by Hall sensors}

\author{J.\ Lefebvre}
\affiliation{Department of Physics, McGill University, Montr\'eal, Canada
H3A 2T8.}
\author{M.\ Hilke}
\affiliation{Department of Physics, McGill University, Montr\'eal, Canada
H3A 2T8.}
\author{Z.\ Altounian}
\affiliation{Department of Physics, McGill University, Montr\'eal, Canada
H3A 2T8.}
\author{K.\ W.\ West}
\affiliation{Bell Laboratories, Alcatel-Lucent, Murray Hill, New Jersey 07974-0636, USA}
\author{L.\ N.\ Pfeiffer}
\affiliation{Bell Laboratories, Alcatel-Lucent, Murray Hill, New Jersey 07974-0636, USA}

\begin{abstract}
We report on magnetization measurements performed on a series
of Fe$_{x}$Ni$_{1-x}$Zr$_{2}$ superconducting metallic glasses with $0\leq x\leq0.5$ using the Hall effect of a nearby 2-dimensional electron gas (2DEG) in a
GaAs/Al$_{0.33}$Ga$_{0.67}$As heterostructure as a local probe. The great sensitivity
of the Hall effect of the 2DEG in such heterostructure is exploited to
determine the magnetization of the superconductor due to the Meissner
effect and flux trapping. The data is used to determine the lower critical field B$_{c1}$ of the superconductors as a function of temperature. Surprisingly large fluctuations in the magnetization are also observed and attributed to the presence of large flux clusters in the superconductor. 

\end{abstract}
\maketitle

Various techniques can be used to obtain information about the
local magnetic profile in type II superconductors. Some, like muon spin
rotation \cite{Lee, Divakar}, neutron scattering \cite{Cubitt, Klein} or
scanning tunneling microscopy \cite{Maggio, TroyanovskiPRL89,
TroyanovskiNat399}, can provide information about the ordering of vortices in
the superconductor without directly probing the magnetic field. Others have
either sufficient spatial resolution or sensitivity to resolve single flux
quanta by directly probing the magnetic field of the vortices. These include for instance,
Lorentz microscopy, magnetic force microscopy, Bitter decoration, scanning
SQUID (superconducting quantum interference device) microscopy, and scanning
Hall probes. Of these, scanning Hall probes offer the best balance between good sensitivity and high spatial resolution \cite{Bending}. Since the beginning of the 1990s,
the Hall resistance of the two-dimensional electron gas (2DEG) that forms at
the interface between GaAs and AlGaAs in GaAs/AlGaAs heterostructures has been
used to probe magnetic flux in superconductors \cite{BendingPRB42, BendingPRL65, OralAPL69, StoddartPRL71}. With a sensitivity of about 3$\times$10$^{-8}$ T$~$Hz$^{-1/2}$
and submicron spatial resolution \cite{Bending}, these can be used to image
the local flux profile in superconductors at low vortex density. 
In the past, Hall probe arrays have been successfully used to image vortices and vortex bundles in high T$_{c}$ superconductors \cite{OralAPL69}. They were also applied to the study of local magnetic profiles \cite{ZeldovPRL73} and their temporal evolution in such superconductors \cite{Karapetrov, Abulafia, Veen}. In these types of experiments, the influence of the inhomogeneous flux profile of the superconductor on the Hall effect of the nearby 2DEG can only be detected if the 2DEG-superconductor separation is very small; this maximal separation is usually approximated as the
distance between vortices, and thus decreases with increasing magnetic field
\cite{Rammer}. Such requirements can be quite stringent, especially in
superconductors having large vortices (large penetration depth $\lambda$) resulting in a magnetic profile inhomogeneity that is rapidly lost upon increase of the external magnetic field.

\begin{figure}
[ptbh]
\begin{center}
\includegraphics[
height=2.8919in,
width=3.4411in
]%
{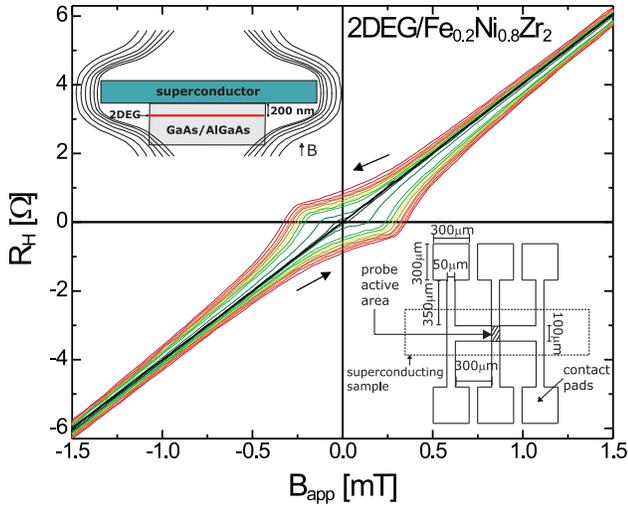}%
\caption{Hall resistance R$_{H}$ of the 2DEG with a sample of superconducting
Fe$_{0.2}$Ni$_{0.8}$Zr$_{2}$ nearby. From larger to smaller hysteresis loop :
T=0.35, 0.53, 0.72, 0.88, 1.11, 1.26, 1.42, 1.57, 1.79, 2.01, 2.31 K. Insets:
(upper) Sketch of experimental geometry showing the bending of magnetic field
lines due to the Meissner effect of the superconductor. (lower) \ Schematic
representation of the Hall bar patterned on the GaAs/AlGaAs.}%
\label{Rh}%
\end{center}
\end{figure}

In this article, we show that using a 2DEG Hall probe with an active area of 100$\times$50 $\mu$m$^{2}$ and at a distance from the
superconductor's surface between 1 and 10 microns, we were able to determine the presence of large vortex clusters in some of our superconductors. A sketch of the geometry proposed for this experiment can be visualized in the upper inset of Fig.\ \ref{Rh}. In an externally applied magnetic field, screening currents are induced in the superconductor due to the Meissner effect. These currents produce the self-field of the superconductor (its magnetization) which fully (or partially) shield the interior of the superconductor from the external field. As a result, the magnetic field threading the nearby 2DEG is composed of the applied field, plus the magnetization of the superconductor. Since the vortices in these metallic glasses are quite large \cite{mythesis, xdep} (close to 1 $\mu$m), no magnetic field inhomogeneity due to these is expected to survive at the 2DEG located over 1 $\mu$m away. In this case, the Hall resistance of the 2DEG reflects the average magnetic field crossing the active area of the 2DEG defined by the Hall junction; the portion of this magnetic field due to the superconductor is attributable to the Meissner effect and gives the local magnetization of the superconductor.  

Performing magnetization measurements using this technique, we could determine the temperature dependence of the lower critical field B$_{c1}$ of the superconductors. In addition, large fluctuations in the local magnetization in some of the superconductors were observed, and found to correlate with the Fe content in the superconductors. We believe that the fluctuations are caused by vortex bundles. A mechanism is proposed to account for the formation of the vortex clusters.

In detail, we use a GaAs/Al$_{0.33}$Ga$_{0.67}$As
heterostructure with a 2DEG sitting 200 nm below the surface with electron
density $n_{e}=1.55\times10^{11}$ cm$^{-2}$ and low-temperature mobility
$\mu=2.86\times10^{6}~$cm$^{2}$V$^{-1}$s$^{-1}$ to measure the low-field
magnetization of a series of metallic glasses Fe$_{x}$Ni$_{1-x}$Zr$_{2}$ with
0 $\leq x\leq$ 0.5. This yields a magnetic field sensitivity of about 4 $\Omega$/mT, independent of size of the Hall probe. On the contrary, the spatial resolution of the Hall probe directly depends on the active area of the 2DEG probe. Therefore, in order to define the active area, a Hall bar pattern, as shown in the lower inset of Fig.\ \ref{Rh}, was scratched on the surface of the heterostructure. The scratching is performed by a diamond tip attached to a fixed
arm; the stage on which the GaAs/AlGaAs sample sits is moved horizontally and
vertically by two motors as controlled by a Labview program which produces the
desired Hall bar pattern. This allows for quick and efficient patterning of 2DEG structures. Indium contacts are deposited in the contact pads of
the Hall bar which is then heated to 400 $
{{}^\circ}%
$C in a sealed quartz tube under vacuum and let to diffuse for 25 minutes in order to contact
the 2DEG below the surface. The superconductor is then placed over the active
area confined by the scratched pattern in the GaAs/AlGaAs and held with vacuum
grease. We find this technique very convenient
because it allows for easy removal and exchange of the superconductor without
chancing degradation of the 2DEG. Also, as it cools, vacuum grease hardens and
contracts and holds the superconductor well in place. The resulting distance
between the 2DEG and superconductor can be estimated from the
capacitance between them; doing this, we obtain this distance to be from 1 to
10 $\mu$m, depending on the sample.

The Fe$_{x}$Ni$_{1-x}$Zr$_{2}$ alloys are prepared by arc-melting appropriate
concentrations of the starting elements Fe (99.9\%), Ni (99.999\%) and Zr
(99.95\%) in Ti-gettered atmosphere. Amorphous ribbons, about 20 $\mu$m thick,
are then obtained from melt-spinning in 40 kPa helium onto a copper wheel
spinning at 50 m/s; the absence of crystallinity was confirmed by the absence
of Bragg peaks in Cu K$_{\alpha}$ x-ray diffraction patterns. 

\begin{figure}
[tbh]
\begin{center}
\includegraphics[
trim=0.000000in 3.862199in 1.484773in 0.000000in,
height=3.5578in,
width=3.4809in
]%
{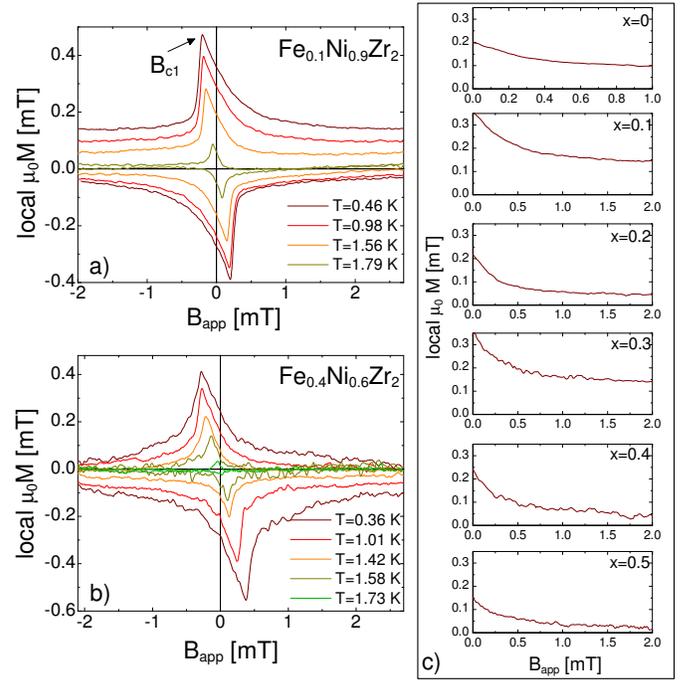}%
\caption{Local magnetization as a function of applied magnetic field for a)
Fe$_{0.1}$Ni$_{0.9}$Zr$_{2}$ and b) Fe$_{0.4}$Ni$_{0.6}$Zr$_{2}$ for different
temperatures and c) only for T$\approx0.35$ K for each x.}%
\label{magnetization}%
\end{center}
\end{figure}

A set of 2DEG Hall resistance R$_{H}$ curves acquired as a function of applied
magnetic field B$_{app}$ on a 2DEG/Fe$_{0.2}$Ni$_{0.8}$Zr$_{2}$ superconductor
composite sample is shown in Fig.\ \ref{Rh}. The different colored curves
correspond to different temperatures as described in the caption. Above the
critical temperature of the superconductor T$_{c}$, the Hall resistance
recovers a linear relationship to B$_{app}$, as it should in the absence of
the superconductor. Below T$_{c}$, the total magnetic field B$_{tot}$
threading the 2DEG is composed of B$_{app}$, plus the demagnetizing field of
the superconductor $\mu_{0}M$, i.e. $\mu_{0}M=B_{tot}-B_{app}$.
B$_{tot}$ is obtained from the measured Hall resistance in the presence of the
superconductor and the conversion from R$_{H}$ to B field is obtained from the Hall constant measured above T$_{c}$
\begin{equation}
B_{tot}=\frac{R_{H}}{\left.  dR_{H}/dB\right\vert _{T>T_{c}}}\text{.}
\end{equation}
The Hall constant is independent of temperature, any dependence of the Hall resistance on temperature is then attributable to the contacts or to a longitudinal contribution. In our case, this longitudinal contribution due to a slight contact misalignment was found to be very small, less than 0.2 $\Omega$ at B=0 and T=0.33 K, and was thus neglected.
The low-field local magnetization loops thus obtained are computed using the linear fit of R$_{H}$ vs B$_{app}$ for T
$>$T$_{c}$ to determine B$_{app}$; results for Fe$_{x}$Ni$_{1-x}$Zr$_{2}$ with x=0.1 and x=0.4 are shown in Fig.\ \ref{magnetization} a) and b). 

\begin{figure}
[ptbh]
\begin{center}
\includegraphics[
trim=0.000000in 5.032627in 0.000000in 0.000000in,
height=2.3583in,
width=3.41in
]%
{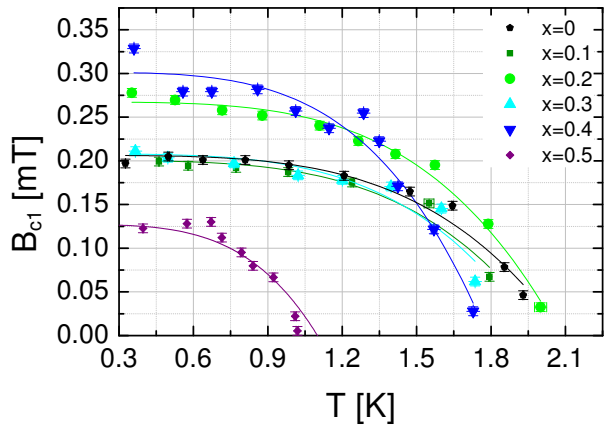}%
\caption{Lower critical field B$_{c1}$ as a function of temperature for
different metallic glasses Fe$_{x}$Ni$_{1-x}$Zr$_{2}$. The lines are fits to
$B_{c1}=B_{c1}\left(  0\right)  (1-\left(  T/T_{c}\right)  ^{4})$ based on a
two-fluid model of superconductivity \cite{Tinkham}.}%
\label{Bc1}%
\end{center}
\end{figure}

We define the location of B$_{c1}$ at the position of the peak in the local magnetization profile, as shown in Fig.\ \ref{Rh}a). B$_{c1}$ values are obtained for the different superconductors Fe$_{x}$Ni$_{1-x}$Zr$_{2}$ with 0 $\leq$ x $\leq$ 0.5 as a function of temperature, Fig.\ \ref{Bc1}. Flux pinning and hysteresis, as well as surface barrier effects often render the observation of B$_{c1}$ difficult, although the weak-pinning properties of these metallic glasses and the local nature of our magnetization measurements make it possible here. Indeed, in strongly-pinned superconductors, the magnetization peak at B$_{c1}$ is often broad and shallow \cite{ShenSST4}; this can be contrasted to the very sharp peak observed here. The local nature of our magnetization measurement is also responsible for this, since averaging is thus only performed over a small portion of the superconductor located in the middle of the superconductor. As a consequence of this, no demagnetizing factor needs to be taken into account for the scaling of the applied magnetic field.

\begin{figure}
[ptbh]
\begin{center}
\includegraphics[
trim=0.000000in 7.441333in 3.711933in 0.000000in,
height=2.4984in,
width=3.4411in
]%
{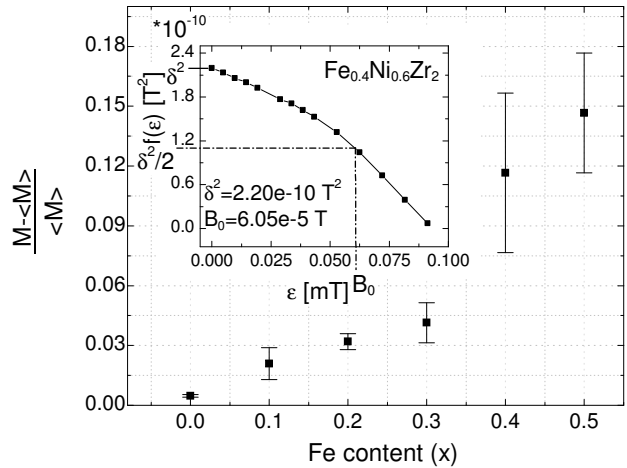}%
\caption{a) Relative fluctuations $\frac{M-\left\langle M\right\rangle
}{\left\langle M\right\rangle }$ as a function of Fe content x in Fe$_{x}%
$Ni$_{1-x}$Zr$_{2}$. Inset: function $f(\varepsilon)$ as described in the text
as a function of magnetic field increment $\varepsilon$, evaluated on
Fe$_{0.4}$Ni$_{0.6}$Zr$_{2}$. }%
\label{noise}%
\end{center}
\end{figure}

\begin{figure}
[ptbh]
\begin{center}
\includegraphics[
trim=0.142984in 8.159068in 5.169240in 0.000000in,
height=3.0415in,
width=3.4402in
]%
{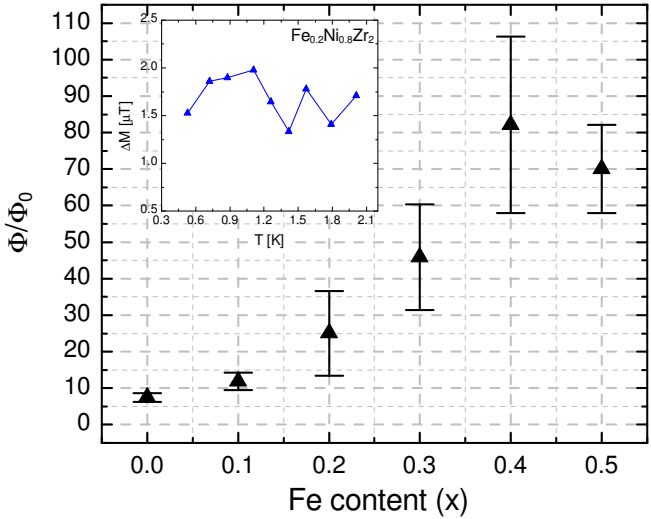}%
\caption{Characteristic magnetic flux of fluctuations as a function of Fe
content\ x in Fe$_{x}$Ni$_{1-x}$Zr$_{2}$. $\Phi/\Phi_{0}$ corresponds to the
number of correlated vortices entering and exiting the area of the superconductor defined by the Hall bar. As
for the data of Fig.\ \ref{noise}, $\Phi/\Phi_{0}$ here is the mean of the
values evaluated over the 4 sections of the magnetization curve (both field
polarities and both sweep directions), and the error bars are the statistical
error on the mean. Inset: Fluctuations $\Delta M=M-\,\left\langle
M\right\rangle $ as a function of temperature.}%
\label{flux}%
\end{center}
\end{figure}

A peculiarity of the magnetization data shown
here is the conspicuous increase in fluctuations in the magnetization signal with
increasing Fe content in the superconductors. This gradual increase in
fluctuations was obtained for superconductors with Fe content x from 0
(smallest fluctuations) to x=0.5 (largest fluctuations). See Fig.\ \ref{magnetization}c) for a portion of the magnetization curve at T $\simeq$ 0.35 K for each superconductor measured to see this evolution. We quantify these
fluctuations in magnetization by computing the relative size of fluctuations
$\frac{M-\left\langle M\right\rangle }{\left\langle M\right\rangle }$ as a
function of Fe content (Fig.\ \ref{noise}). $\left\langle M\right\rangle $ is
obtained from fitting a 4$^{th}$ order polynomial to the magnetization signal
as necessary to determine the mean of a non-constant signal. This is computed over the four
sections of the magnetization curve corresponding to both polarities of the
applied field and both sweep directions; the values shown in Fig.\ \ref{noise}
are the mean of these four evaluations for the lowest temperature probed, and the error bars the statistical
error. The size of fluctuations is observed to be pretty constant with temperature for the six samples measured (see inset of Fig.\ \ref{flux} for x=0.2); no particular trend is observed in $\Delta M=M-\,\left\langle
M\right\rangle $ as a function of temperature.

Since the magnetization represents a sum over magnetic moments $M=
{\displaystyle\sum\nolimits_{i}^{N}}
m_{i}$, M depends on both the total number $N=V/a^{3}$ of these moments,
and on their magnitude $m_{i}$ ($V$ is the volume of the superconductor, and
$a^{3}$ is the characteristic size of grains). Accordingly, from the
magnetization fluctuations, one can determine the minimum value for the
size of grains and the characteristic magnetic field $B_{0}$ of fluctuations.
More precisely, since $\frac{M-\left\langle M\right\rangle
}{\left\langle M\right\rangle } < \frac{1}{\sqrt{N}}$, assuming independent and maximally fluctuating grains, we obtain $a > 13~\mu m$ for the x=0.5
alloy. Furthermore, the characteristic field of fluctuations $B_{0}$ can be
estimated from computation of the auto-covariance function $f\left(
\varepsilon\right)  =\left\langle M(B)M(B+\varepsilon)\right\rangle
-\left\langle M(B)\right\rangle ^{2}$, where $\varepsilon$ is a small magnetic
field increment, and $\sqrt{f\left(  0\right)  }=\delta=\sqrt{\left\langle
M(B)^{2}\right\rangle -\left\langle M(B)\right\rangle ^{2}}$ is the usual
standard deviation expression; $B_{0}$ is the value of $\varepsilon$ at which
$f\left(  \varepsilon\right)  =\delta^{2}/2$ (inset of Fig.\ \ref{noise}). As
expected, $f\left(  \varepsilon\right)  $ decreases with increasing
$\varepsilon$ as correlations diminish. The characteristic magnetic flux of
fluctuations can then be computed as $\Phi=B_{0}\times A$ where $A$ is the active area of the Hall probe perpendicular to the field (Fig.\ \ref{flux}), from
which it can be deduced that the fluctuations arise due to the entry and exit of
vortex bundles in and out of the area of the superconductor defined by the Hall probe. In the samples containing the
largest amount of Fe, these correlated flux movements can be quite large with 70 to 80 vortices. These processes are visible because our Hall probe provides a local measurement of the magnetization and averaging of the signal is performed over only a small part of the superconductor, such effects are typically not visible in global magnetization measurements.

The occurrence of vortex movement in
such large correlated bundles and its dependence on Fe content x in the superconductors can be understood as follows: The superconductor is composed of two different phases, one having weaker pinning properties than
the other, such that vortex entry and exit would be privileged there. The
number density and size of inhomogeneities would then increase with Fe content
to yield larger fluctuations induced by larger flux bundles. Such a scenario
would be consistent with the possibility that our superconducting metallic
glasses are composed of Ni-rich and Fe-rich clusters having different short
range order (SRO). Since Fe and Ni atoms have very similar sizes, no change in
geometrical short range order (GSRO) is expected upon substitution of Ni for
Fe in these alloys. This is however not necessarily true of the chemical short
range order (CSRO); pertaining mainly to the atomic species of
nearest-neighbors, CSRO could vary because Fe and Ni do not have the same
electronic structure. This question is especially relevant in these metallic
glasses because it is known that the first crystallization products of
NiZr$_{2}$ assume a body-centered-tetragonal (bct) structure while those of
FeZr$_{2}$ are face-centered-cubic (fcc), and because the atomic arrangement in the
amorphous phase is assumed to be very close to that of the first
crystallization products \cite{Wang, McKamey}, a transition in SRO with $x$ in amorphous Fe$_{x}%
$Ni$_{1-x}$Zr$_{2}$ could be expected. According to this picture, a phase
having SRO resembling the fcc arrangement of FeZr$_{2}$ could develop with
increasing Fe content in the alloys; most probably having different pinning
properties, it could lead to vortex clustering. 

Other evidences published
elsewhere point to two-phase superconductivity in the alloys with large Fe
content x=0.5 and 0.6, namely a normal-to-superconducting state transition
showing a double step, and anomalous clockwise hysteresis at the B$_{c2}$
transition \cite{xdep, mythesis}. If the magnetization fluctuations observed here are indeed the hallmark of a two-phase superconductor, it could lead to the interesting conclusion that a structural transition exists in these amorphous alloys, although it has not been seen before despite investigation, for instance using
M\"{o}ssbauer spectroscopy \cite{Dikeakos}. Such small differences in the atomic ordering in the amorphous state cannot be highlighted easily due to the small scale involved, but for this to take place, it means that using the small vortex core ($\xi\simeq7~$nm in these metallic glasses \cite{HilkePRL91, xdep}) as a probe, it is possible to resolve such small changes. Considering the dramatic increase in the size of fluctuations
for Fe content $0.4\leq x\leq0.5$, if a structural transition exists in these alloys, it must be located close to x=0.4, but with a gradual increase of an Fe-rich phase starting with $x=0.1$.

However, it is not clear how structural inhomogeneities could
lead to magnetization fluctuations as a function of magnetic field. It could
be the case if the respective sizes of the Ni-rich and Fe-rich regions, or the
boundary between these two phases, changed as a function of field, but this
should be invariant. This picture is however consistent with the observation
that $\Delta M$ does not have a clear dependence on temperature
(Fig.\ \ref{flux}, inset), meaning that the fluctuations are issued from a
phenomenon constant with respect to that variable, and not from a
thermally-dependent process.

In addition, the Fe content modifies the pinning properties, which are extremely weak in these amorphous
alloys. Indeed, as discussed elsewhere \cite{HilkePRL91, LefebvrePRB74, xdep}, pinning in
these alloys is from 10 to 1000 times weaker than in other similar
superconductors \cite{BerlincourtPRL6, KesPRB28, WordenweberPRB33,
GeersPRB63}, with a critical current density J$_{c}<3$ A/cm$^{2}$ at B=0.15B$_{c2}$. Also, it has been shown that the pinning force decreases by
almost a factor of 5 when going from the alloys with a low Fe content (x=0 and
0.1) to the alloys with a high Fe content (x=0.5 and 0.6) \cite{xdep}. This is
partly due to the larger size of vortices ($\lambda$ and $\xi_{GL}$ almost
double from x=0 to x=0.6) in alloys with a high Fe content \cite{xdep,
mythesis}, which confers less efficient pinning for pinning sites of similar
size. In addition, the larger size of vortices favors vortex-vortex
interactions, thus enhancing collective effects, which also contributes to
diminishing effective pinning. In these superconductors, it was shown that
vortices arrange in the Bragg glass (BG) phase at driving forces lower than
the depinning threshold and at low vortex density \cite{LefebvrePRB78}, such
that even at such low magnetic field ($\lesssim2$ mT) collective effects are
significant; the existence of elastic interactions between vortices is a
necessary condition for the BG phase \cite{Giamarchi}. Therefore, vortices will likely form correlated clusters of increasing size with Fe content, which will induce fluctuations in the magnetization as
bundles enter and exit the boundary defined by the Hall bar on the superconductor. 

To summarize, we have shown that using our Hall probe, we could measure the low-field magnetization of a series of Fe$_{x}$Ni$_{1-x}$Zr$_{2}$ metallic glass superconductors and the dependence of B$_{c1}$ on temperature. We further observed fluctuations in the magnetization signal, the magnitude of which increases with the Fe content in the superconductors. The origin of these fluctuations is linked to the presence of large vortex bundles in the superconductors. These fluctuations are consistent with an increasingly two-phase system (Fe-rich and Ni-rich) with different SRO and pinning properties leading to an increased effective vortex-vortex interaction which enhances collective vortex movement.


\begin{thebibliography}{32}
\expandafter\ifx\csname natexlab\endcsname\relax\def\natexlab#1{#1}\fi
\expandafter\ifx\csname bibnamefont\endcsname\relax
  \def\bibnamefont#1{#1}\fi
\expandafter\ifx\csname bibfnamefont\endcsname\relax
  \def\bibfnamefont#1{#1}\fi
\expandafter\ifx\csname citenamefont\endcsname\relax
  \def\citenamefont#1{#1}\fi
\expandafter\ifx\csname url\endcsname\relax
  \def\url#1{\texttt{#1}}\fi
\expandafter\ifx\csname urlprefix\endcsname\relax\def\urlprefix{URL }\fi
\providecommand{\bibinfo}[2]{#2}
\providecommand{\eprint}[2][]{\url{#2}}

\bibitem[{\citenamefont{Lee et~al.}(1993)\citenamefont{Lee, Zimmermann, Keller,
  Warden, Savi\ifmmode~\acute{c}\else \'{c}\fi{}, Schauwecker, Zech, Cubitt,
  Forgan, Kes et~al.}}]{Lee}
\bibinfo{author}{\bibfnamefont{S.~L.} \bibnamefont{Lee}},
  \bibinfo{author}{\bibfnamefont{P.}~\bibnamefont{Zimmermann}},
  \bibinfo{author}{\bibfnamefont{H.}~\bibnamefont{Keller}},
  \bibinfo{author}{\bibfnamefont{M.}~\bibnamefont{Warden}},
  \bibinfo{author}{\bibfnamefont{I.~M.}
  \bibnamefont{Savi\ifmmode~\acute{c}\else \'{c}\fi{}}},
  \bibinfo{author}{\bibfnamefont{R.}~\bibnamefont{Schauwecker}},
  \bibinfo{author}{\bibfnamefont{D.}~\bibnamefont{Zech}},
  \bibinfo{author}{\bibfnamefont{R.}~\bibnamefont{Cubitt}},
  \bibinfo{author}{\bibfnamefont{E.~M.} \bibnamefont{Forgan}},
  \bibinfo{author}{\bibfnamefont{P.~H.} \bibnamefont{Kes}},
  \bibnamefont{et~al.}, \bibinfo{journal}{Phys. Rev. Lett.}
  \textbf{\bibinfo{volume}{71}}, \bibinfo{pages}{3862} (\bibinfo{year}{1993}).

\bibitem[{\citenamefont{Divakar et~al.}(2004)\citenamefont{Divakar, Drew, Lee,
  Gilardi, Mesot, Ogrin, Charalambous, Forgan, Menon, Momono et~al.}}]{Divakar}
\bibinfo{author}{\bibfnamefont{U.}~\bibnamefont{Divakar}},
  \bibinfo{author}{\bibfnamefont{A.~J.} \bibnamefont{Drew}},
  \bibinfo{author}{\bibfnamefont{S.~L.} \bibnamefont{Lee}},
  \bibinfo{author}{\bibfnamefont{R.}~\bibnamefont{Gilardi}},
  \bibinfo{author}{\bibfnamefont{J.}~\bibnamefont{Mesot}},
  \bibinfo{author}{\bibfnamefont{F.~Y.} \bibnamefont{Ogrin}},
  \bibinfo{author}{\bibfnamefont{D.}~\bibnamefont{Charalambous}},
  \bibinfo{author}{\bibfnamefont{E.~M.} \bibnamefont{Forgan}},
  \bibinfo{author}{\bibfnamefont{G.~I.} \bibnamefont{Menon}},
  \bibinfo{author}{\bibfnamefont{N.}~\bibnamefont{Momono}},
  \bibnamefont{et~al.}, \bibinfo{journal}{Phys. Rev. Lett.}
  \textbf{\bibinfo{volume}{92}}, \bibinfo{pages}{237004}
  (\bibinfo{year}{2004}).

\bibitem[{\citenamefont{Cubitt et~al.}(1993)\citenamefont{Cubitt, Forgan, Yang,
  Lee, Paul, Mook, Yethiraj, Kes, Li, Menovsky et~al.}}]{Cubitt}
\bibinfo{author}{\bibfnamefont{R.}~\bibnamefont{Cubitt}},
  \bibinfo{author}{\bibfnamefont{E.~M.} \bibnamefont{Forgan}},
  \bibinfo{author}{\bibfnamefont{G.}~\bibnamefont{Yang}},
  \bibinfo{author}{\bibfnamefont{S.~L.} \bibnamefont{Lee}},
  \bibinfo{author}{\bibfnamefont{D.~M.} \bibnamefont{Paul}},
  \bibinfo{author}{\bibfnamefont{H.~A.} \bibnamefont{Mook}},
  \bibinfo{author}{\bibfnamefont{M.}~\bibnamefont{Yethiraj}},
  \bibinfo{author}{\bibfnamefont{P.~H.} \bibnamefont{Kes}},
  \bibinfo{author}{\bibfnamefont{T.~W.} \bibnamefont{Li}},
  \bibinfo{author}{\bibfnamefont{A.~A.} \bibnamefont{Menovsky}},
  \bibnamefont{et~al.}, \bibinfo{journal}{Nature}
  \textbf{\bibinfo{volume}{365}}, \bibinfo{pages}{407} (\bibinfo{year}{1993}).

\bibitem[{\citenamefont{Klein et~al.}(2001)\citenamefont{Klein, Joumard,
  Blanchard, Marcus, Cubitt, Giamarchi, and Doussal}}]{Klein}
\bibinfo{author}{\bibfnamefont{T.}~\bibnamefont{Klein}},
  \bibinfo{author}{\bibfnamefont{I.}~\bibnamefont{Joumard}},
  \bibinfo{author}{\bibfnamefont{S.}~\bibnamefont{Blanchard}},
  \bibinfo{author}{\bibfnamefont{J.}~\bibnamefont{Marcus}},
  \bibinfo{author}{\bibfnamefont{R.}~\bibnamefont{Cubitt}},
  \bibinfo{author}{\bibfnamefont{T.}~\bibnamefont{Giamarchi}},
  \bibnamefont{and} \bibinfo{author}{\bibfnamefont{P.~L.}
  \bibnamefont{Doussal}}, \bibinfo{journal}{Nature}
  \textbf{\bibinfo{volume}{413}}, \bibinfo{pages}{404} (\bibinfo{year}{2001}).

\bibitem[{\citenamefont{Maggio-Aprile et~al.}(1995)\citenamefont{Maggio-Aprile,
  Renner, Erb, Walker, and Fischer}}]{Maggio}
\bibinfo{author}{\bibfnamefont{I.}~\bibnamefont{Maggio-Aprile}},
  \bibinfo{author}{\bibfnamefont{C.}~\bibnamefont{Renner}},
  \bibinfo{author}{\bibfnamefont{A.}~\bibnamefont{Erb}},
  \bibinfo{author}{\bibfnamefont{E.}~\bibnamefont{Walker}}, \bibnamefont{and}
  \bibinfo{author}{\bibfnamefont{O.}~\bibnamefont{Fischer}},
  \bibinfo{journal}{Phys. Rev. Lett.} \textbf{\bibinfo{volume}{75}},
  \bibinfo{pages}{2754} (\bibinfo{year}{1995}).

\bibitem[{\citenamefont{Troyanovski et~al.}(2002)\citenamefont{Troyanovski, van
  Hecke, Saha, Aarts, and Kes}}]{TroyanovskiPRL89}
\bibinfo{author}{\bibfnamefont{A.~M.} \bibnamefont{Troyanovski}},
  \bibinfo{author}{\bibfnamefont{M.}~\bibnamefont{van Hecke}},
  \bibinfo{author}{\bibfnamefont{N.}~\bibnamefont{Saha}},
  \bibinfo{author}{\bibfnamefont{J.}~\bibnamefont{Aarts}}, \bibnamefont{and}
  \bibinfo{author}{\bibfnamefont{P.~H.} \bibnamefont{Kes}},
  \bibinfo{journal}{Phys. Rev. Lett.} \textbf{\bibinfo{volume}{89}},
  \bibinfo{pages}{147006} (\bibinfo{year}{2002}).

\bibitem[{\citenamefont{Troyanovski et~al.}(1999)\citenamefont{Troyanovski,
  Aarts, and Kes}}]{TroyanovskiNat399}
\bibinfo{author}{\bibfnamefont{A.~M.} \bibnamefont{Troyanovski}},
  \bibinfo{author}{\bibfnamefont{J.}~\bibnamefont{Aarts}}, \bibnamefont{and}
  \bibinfo{author}{\bibfnamefont{P.~H.} \bibnamefont{Kes}},
  \bibinfo{journal}{Nature} \textbf{\bibinfo{volume}{399}},
  \bibinfo{pages}{665} (\bibinfo{year}{1999}).

\bibitem[{\citenamefont{Bending}(1999)}]{Bending}
\bibinfo{author}{\bibfnamefont{S.~J.} \bibnamefont{Bending}},
  \bibinfo{journal}{Adv. Phys.} \textbf{\bibinfo{volume}{48}},
  \bibinfo{pages}{449} (\bibinfo{year}{1999}).

\bibitem[{\citenamefont{Bending
  et~al.}(1990{\natexlab{a}})\citenamefont{Bending, von Klitzing, and
  Ploog}}]{BendingPRB42}
\bibinfo{author}{\bibfnamefont{S.~J.} \bibnamefont{Bending}},
  \bibinfo{author}{\bibfnamefont{K.}~\bibnamefont{von Klitzing}},
  \bibnamefont{and} \bibinfo{author}{\bibfnamefont{K.}~\bibnamefont{Ploog}},
  \bibinfo{journal}{Phys. Rev. B} \textbf{\bibinfo{volume}{42}},
  \bibinfo{pages}{9859} (\bibinfo{year}{1990}{\natexlab{a}}).

\bibitem[{\citenamefont{Bending
  et~al.}(1990{\natexlab{b}})\citenamefont{Bending, von Klitzing, and
  Ploog}}]{BendingPRL65}
\bibinfo{author}{\bibfnamefont{S.~J.} \bibnamefont{Bending}},
  \bibinfo{author}{\bibfnamefont{K.}~\bibnamefont{von Klitzing}},
  \bibnamefont{and} \bibinfo{author}{\bibfnamefont{K.}~\bibnamefont{Ploog}},
  \bibinfo{journal}{Phys. Rev. Lett.} \textbf{\bibinfo{volume}{65}},
  \bibinfo{pages}{1060} (\bibinfo{year}{1990}{\natexlab{b}}).

\bibitem[{\citenamefont{Oral et~al.}(1996)\citenamefont{Oral, Bending, and
  Henini}}]{OralAPL69}
\bibinfo{author}{\bibfnamefont{A.}~\bibnamefont{Oral}},
  \bibinfo{author}{\bibfnamefont{S.~J.} \bibnamefont{Bending}},
  \bibnamefont{and} \bibinfo{author}{\bibfnamefont{M.}~\bibnamefont{Henini}},
  \bibinfo{journal}{Applied Physics Letters} \textbf{\bibinfo{volume}{69}},
  \bibinfo{pages}{1324} (\bibinfo{year}{1996}).

\bibitem[{\citenamefont{Stoddart et~al.}(1993)\citenamefont{Stoddart, Bending,
  Geim, and Henini}}]{StoddartPRL71}
\bibinfo{author}{\bibfnamefont{S.~T.} \bibnamefont{Stoddart}},
  \bibinfo{author}{\bibfnamefont{S.~J.} \bibnamefont{Bending}},
  \bibinfo{author}{\bibfnamefont{A.~K.} \bibnamefont{Geim}}, \bibnamefont{and}
  \bibinfo{author}{\bibfnamefont{M.}~\bibnamefont{Henini}},
  \bibinfo{journal}{Phys. Rev. Lett.} \textbf{\bibinfo{volume}{71}},
  \bibinfo{pages}{3854} (\bibinfo{year}{1993}).

\bibitem[{\citenamefont{Zeldov et~al.}(1994)\citenamefont{Zeldov, Larkin,
  Geshkenbein, Konczykowski, Majer, Khaykovich, Vinokur, and
  Shtrikman}}]{ZeldovPRL73}
\bibinfo{author}{\bibfnamefont{E.}~\bibnamefont{Zeldov}},
  \bibinfo{author}{\bibfnamefont{A.~I.} \bibnamefont{Larkin}},
  \bibinfo{author}{\bibfnamefont{V.~B.} \bibnamefont{Geshkenbein}},
  \bibinfo{author}{\bibfnamefont{M.}~\bibnamefont{Konczykowski}},
  \bibinfo{author}{\bibfnamefont{D.}~\bibnamefont{Majer}},
  \bibinfo{author}{\bibfnamefont{B.}~\bibnamefont{Khaykovich}},
  \bibinfo{author}{\bibfnamefont{V.~M.} \bibnamefont{Vinokur}},
  \bibnamefont{and}
  \bibinfo{author}{\bibfnamefont{H.}~\bibnamefont{Shtrikman}},
  \bibinfo{journal}{Phys. Rev. Lett.} \textbf{\bibinfo{volume}{73}},
  \bibinfo{pages}{1428} (\bibinfo{year}{1994}).

\bibitem[{\citenamefont{Karapetrov et~al.}(1999)\citenamefont{Karapetrov,
  Cambel, Kwok, Nikolova, Crabtree, Zheng, and Veal}}]{Karapetrov}
\bibinfo{author}{\bibfnamefont{G.}~\bibnamefont{Karapetrov}},
  \bibinfo{author}{\bibfnamefont{V.}~\bibnamefont{Cambel}},
  \bibinfo{author}{\bibfnamefont{W.~K.} \bibnamefont{Kwok}},
  \bibinfo{author}{\bibfnamefont{R.}~\bibnamefont{Nikolova}},
  \bibinfo{author}{\bibfnamefont{G.~W.} \bibnamefont{Crabtree}},
  \bibinfo{author}{\bibfnamefont{H.}~\bibnamefont{Zheng}}, \bibnamefont{and}
  \bibinfo{author}{\bibfnamefont{B.~W.} \bibnamefont{Veal}},
  \bibinfo{journal}{Journal of Applied Physics} \textbf{\bibinfo{volume}{86}},
  \bibinfo{pages}{6282} (\bibinfo{year}{1999}).

\bibitem[{\citenamefont{Abulafia et~al.}(1995)\citenamefont{Abulafia, Shaulov,
  Wolfus, Prozorov, Burlachkov, Yeshurun, Majer, Zeldov, and
  Vinokur}}]{Abulafia}
\bibinfo{author}{\bibfnamefont{Y.}~\bibnamefont{Abulafia}},
  \bibinfo{author}{\bibfnamefont{A.}~\bibnamefont{Shaulov}},
  \bibinfo{author}{\bibfnamefont{Y.}~\bibnamefont{Wolfus}},
  \bibinfo{author}{\bibfnamefont{R.}~\bibnamefont{Prozorov}},
  \bibinfo{author}{\bibfnamefont{L.}~\bibnamefont{Burlachkov}},
  \bibinfo{author}{\bibfnamefont{Y.}~\bibnamefont{Yeshurun}},
  \bibinfo{author}{\bibfnamefont{D.}~\bibnamefont{Majer}},
  \bibinfo{author}{\bibfnamefont{E.}~\bibnamefont{Zeldov}}, \bibnamefont{and}
  \bibinfo{author}{\bibfnamefont{V.~M.} \bibnamefont{Vinokur}},
  \bibinfo{journal}{Phys. Rev. Lett.} \textbf{\bibinfo{volume}{75}},
  \bibinfo{pages}{2404} (\bibinfo{year}{1995}).

\bibitem[{\citenamefont{van Veen et~al.}(1999)\citenamefont{van Veen,
  Verbruggen, van~der Drift, Radelaar, Anders, and Jaeger}}]{Veen}
\bibinfo{author}{\bibfnamefont{R.~G.} \bibnamefont{van Veen}},
  \bibinfo{author}{\bibfnamefont{A.~H.} \bibnamefont{Verbruggen}},
  \bibinfo{author}{\bibfnamefont{E.}~\bibnamefont{van~der Drift}},
  \bibinfo{author}{\bibfnamefont{S.}~\bibnamefont{Radelaar}},
  \bibinfo{author}{\bibfnamefont{S.}~\bibnamefont{Anders}}, \bibnamefont{and}
  \bibinfo{author}{\bibfnamefont{H.~M.} \bibnamefont{Jaeger}},
  \bibinfo{journal}{Review of Scientific Instruments}
  \textbf{\bibinfo{volume}{70}}, \bibinfo{pages}{1767} (\bibinfo{year}{1999}).

\bibitem[{\citenamefont{Rammer and Shelankov}(1987)}]{Rammer}
\bibinfo{author}{\bibfnamefont{J.}~\bibnamefont{Rammer}} \bibnamefont{and}
  \bibinfo{author}{\bibfnamefont{A.~L.} \bibnamefont{Shelankov}},
  \bibinfo{journal}{Phys. Rev. B} \textbf{\bibinfo{volume}{36}},
  \bibinfo{pages}{3135} (\bibinfo{year}{1987}).

\bibitem[{\citenamefont{Lefebvre}(2008)}]{mythesis}
\bibinfo{author}{\bibfnamefont{J.}~\bibnamefont{Lefebvre}}, Ph.D. thesis,
  \bibinfo{address}{McGill University, {Montr\'{e}al}, Canada}
  (\bibinfo{year}{2008}).

\bibitem[{\citenamefont{Lefebvre
  et~al.}(2008{\natexlab{a}})\citenamefont{Lefebvre, Hilke, and
  Altounian}}]{xdep}
\bibinfo{author}{\bibfnamefont{J.}~\bibnamefont{Lefebvre}},
  \bibinfo{author}{\bibfnamefont{M.}~\bibnamefont{Hilke}}, \bibnamefont{and}
  \bibinfo{author}{\bibfnamefont{Z.}~\bibnamefont{Altounian}}
  (\bibinfo{year}{2008}{\natexlab{a}}), \bibinfo{note}{to be published}.

\bibitem[{\citenamefont{Tinkham}(1996)}]{Tinkham}
\bibinfo{author}{\bibfnamefont{M.}~\bibnamefont{Tinkham}},
  \emph{\bibinfo{title}{Introduction to Superconductivity}}
  (\bibinfo{publisher}{McGraw-Hill}, \bibinfo{year}{1996}).

\bibitem[{\citenamefont{Shen et~al.}(1991)\citenamefont{Shen, Elliott, Cornish,
  Williams, Westwood, Lin, Liang, Richie, Frost, and Jones}}]{ShenSST4}
\bibinfo{author}{\bibfnamefont{T.~H.} \bibnamefont{Shen}},
  \bibinfo{author}{\bibfnamefont{M.}~\bibnamefont{Elliott}},
  \bibinfo{author}{\bibfnamefont{A.~E.} \bibnamefont{Cornish}},
  \bibinfo{author}{\bibfnamefont{R.~H.} \bibnamefont{Williams}},
  \bibinfo{author}{\bibfnamefont{D.}~\bibnamefont{Westwood}},
  \bibinfo{author}{\bibfnamefont{C.~T.} \bibnamefont{Lin}},
  \bibinfo{author}{\bibfnamefont{W.~Y.} \bibnamefont{Liang}},
  \bibinfo{author}{\bibfnamefont{D.~A.} \bibnamefont{Richie}},
  \bibinfo{author}{\bibfnamefont{J.~E.~F.} \bibnamefont{Frost}},
  \bibnamefont{and} \bibinfo{author}{\bibfnamefont{G.~A.~C.}
  \bibnamefont{Jones}}, \bibinfo{journal}{Superconductor Science and
  Technology} \textbf{\bibinfo{volume}{4}}, \bibinfo{pages}{232}
  (\bibinfo{year}{1991}).

\bibitem[{\citenamefont{Wang}(1979)}]{Wang}
\bibinfo{author}{\bibfnamefont{R.}~\bibnamefont{Wang}},
  \bibinfo{journal}{Nature} \textbf{\bibinfo{volume}{278}},
  \bibinfo{pages}{700} (\bibinfo{year}{1979}).

\bibitem[{\citenamefont{McKamey et~al.}(1986)\citenamefont{McKamey, Kroeger,
  Easton, and Scarbrough}}]{McKamey}
\bibinfo{author}{\bibfnamefont{C.}~\bibnamefont{McKamey}},
  \bibinfo{author}{\bibfnamefont{D.~M.} \bibnamefont{Kroeger}},
  \bibinfo{author}{\bibfnamefont{D.~S.} \bibnamefont{Easton}},
  \bibnamefont{and} \bibinfo{author}{\bibfnamefont{J.~O.}
  \bibnamefont{Scarbrough}}, \bibinfo{journal}{J. Mater. Sci.}
  \textbf{\bibinfo{volume}{21}}, \bibinfo{pages}{3863} (\bibinfo{year}{1986}).

\bibitem[{\citenamefont{Dikeakos et~al.}(1999)\citenamefont{Dikeakos,
  Altounian, Ryan, and Kwon}}]{Dikeakos}
\bibinfo{author}{\bibfnamefont{M.}~\bibnamefont{Dikeakos}},
  \bibinfo{author}{\bibfnamefont{Z.}~\bibnamefont{Altounian}},
  \bibinfo{author}{\bibfnamefont{D.~H.} \bibnamefont{Ryan}}, \bibnamefont{and}
  \bibinfo{author}{\bibfnamefont{S.~J.} \bibnamefont{Kwon}},
  \bibinfo{journal}{J. Non-Cryst. Solids} \textbf{\bibinfo{volume}{250-252}},
  \bibinfo{pages}{637} (\bibinfo{year}{1999}).

\bibitem[{\citenamefont{Hilke et~al.}(2003)\citenamefont{Hilke, Reid, Gagnon,
  and Altounian}}]{HilkePRL91}
\bibinfo{author}{\bibfnamefont{M.}~\bibnamefont{Hilke}},
  \bibinfo{author}{\bibfnamefont{S.}~\bibnamefont{Reid}},
  \bibinfo{author}{\bibfnamefont{R.}~\bibnamefont{Gagnon}}, \bibnamefont{and}
  \bibinfo{author}{\bibfnamefont{Z.}~\bibnamefont{Altounian}},
  \bibinfo{journal}{Phys. Rev. Lett.} \textbf{\bibinfo{volume}{91}},
  \bibinfo{pages}{127004} (\bibinfo{year}{2003}).

\bibitem[{\citenamefont{Lefebvre et~al.}(2006)\citenamefont{Lefebvre, Hilke,
  Gagnon, and Altounian}}]{LefebvrePRB74}
\bibinfo{author}{\bibfnamefont{J.}~\bibnamefont{Lefebvre}},
  \bibinfo{author}{\bibfnamefont{M.}~\bibnamefont{Hilke}},
  \bibinfo{author}{\bibfnamefont{R.}~\bibnamefont{Gagnon}}, \bibnamefont{and}
  \bibinfo{author}{\bibfnamefont{Z.}~\bibnamefont{Altounian}},
  \bibinfo{journal}{Phys. Rev. B} \textbf{\bibinfo{volume}{74}},
  \bibinfo{eid}{174509} (\bibinfo{year}{2006}).

\bibitem[{\citenamefont{Berlincourt et~al.}(1961)\citenamefont{Berlincourt,
  Hake, and Leslie}}]{BerlincourtPRL6}
\bibinfo{author}{\bibfnamefont{T.~G.} \bibnamefont{Berlincourt}},
  \bibinfo{author}{\bibfnamefont{R.~R.} \bibnamefont{Hake}}, \bibnamefont{and}
  \bibinfo{author}{\bibfnamefont{D.~H.} \bibnamefont{Leslie}},
  \bibinfo{journal}{Phys. Rev. Lett.} \textbf{\bibinfo{volume}{6}},
  \bibinfo{pages}{671} (\bibinfo{year}{1961}).

\bibitem[{\citenamefont{Kes and Tsuei}(1983)}]{KesPRB28}
\bibinfo{author}{\bibfnamefont{P.~H.} \bibnamefont{Kes}} \bibnamefont{and}
  \bibinfo{author}{\bibfnamefont{C.~C.} \bibnamefont{Tsuei}},
  \bibinfo{journal}{Phys. Rev. B} \textbf{\bibinfo{volume}{28}},
  \bibinfo{pages}{5126} (\bibinfo{year}{1983}).

\bibitem[{\citenamefont{W\"ordenweber et~al.}(1986)\citenamefont{W\"ordenweber,
  Kes, and Tsuei}}]{WordenweberPRB33}
\bibinfo{author}{\bibfnamefont{R.}~\bibnamefont{W\"ordenweber}},
  \bibinfo{author}{\bibfnamefont{P.~H.} \bibnamefont{Kes}}, \bibnamefont{and}
  \bibinfo{author}{\bibfnamefont{C.~C.} \bibnamefont{Tsuei}},
  \bibinfo{journal}{Phys. Rev. B} \textbf{\bibinfo{volume}{33}},
  \bibinfo{pages}{3172} (\bibinfo{year}{1986}).

\bibitem[{\citenamefont{Geers et~al.}(2001)\citenamefont{Geers, Attanasio,
  Hesselberth, Aarts, and Kes}}]{GeersPRB63}
\bibinfo{author}{\bibfnamefont{J.~M.~E.} \bibnamefont{Geers}},
  \bibinfo{author}{\bibfnamefont{C.}~\bibnamefont{Attanasio}},
  \bibinfo{author}{\bibfnamefont{M.~B.~S.} \bibnamefont{Hesselberth}},
  \bibinfo{author}{\bibfnamefont{J.}~\bibnamefont{Aarts}}, \bibnamefont{and}
  \bibinfo{author}{\bibfnamefont{P.~H.} \bibnamefont{Kes}},
  \bibinfo{journal}{Phys. Rev. B} \textbf{\bibinfo{volume}{63}},
  \bibinfo{pages}{094511} (\bibinfo{year}{2001}).

\bibitem[{\citenamefont{Lefebvre
  et~al.}(2008{\natexlab{b}})\citenamefont{Lefebvre, Hilke, and
  Altounian}}]{LefebvrePRB78}
\bibinfo{author}{\bibfnamefont{J.}~\bibnamefont{Lefebvre}},
  \bibinfo{author}{\bibfnamefont{M.}~\bibnamefont{Hilke}}, \bibnamefont{and}
  \bibinfo{author}{\bibfnamefont{Z.}~\bibnamefont{Altounian}},
  \bibinfo{journal}{Phys. Rev. B} \textbf{\bibinfo{volume}{78}},
  \bibinfo{eid}{134506} (\bibinfo{year}{2008}{\natexlab{b}}).

\bibitem[{\citenamefont{Giamarchi and Bhattacharya}(2001)}]{Giamarchi}
\bibinfo{author}{\bibfnamefont{T.}~\bibnamefont{Giamarchi}} \bibnamefont{and}
  \bibinfo{author}{\bibfnamefont{S.}~\bibnamefont{Bhattacharya}},
  \emph{\bibinfo{title}{Vortex phases}} (\bibinfo{year}{2001}),
  \eprint{cond-mat/0111052v1}.

\end{thebibliography}
\end{document}